\documentclass[twocolumn,american,superscriptaddress,showpacs]{revtex4-1}
\usepackage{color}
\usepackage{babel}
\usepackage{float}
\usepackage{graphicx}
\usepackage[unicode=true,pdfusetitle,
 bookmarks=true,bookmarksnumbered=true,bookmarksopen=true,bookmarksopenlevel=1,
 breaklinks=true,pdfborder={0 0 1},backref=false,colorlinks=true]
 {hyperref}
\hypersetup{
 linkcolor=red, citecolor=black, urlcolor=blue, filecolor=blue, pdfpagelayout=OneColumn, pdfnewwindow=true, pdfstartview=XYZ, plainpages=false}

\makeatletter

\def\jb{\color{black}}

\makeatother

\begin{document}
\author{K. Koepernik}
\affiliation{IFW Dresden, P.O. Box 270116, 01171 Dresden, Germany}
\author{D. Kasinathan}
\affiliation{Max-Planck-Institut f{\"u}r Chemische Physik fester Stoffe, N{\"o}thnitzer Strasse 40, 01187 Dresden, Germany}
\author{D.V. Efremov}
\affiliation{IFW Dresden, P.O. Box 270116, 01171 Dresden, Germany}
\author{Seunghyun Khim}
\affiliation{IFW Dresden, P.O. Box 270116, 01171 Dresden, Germany}
\author{Sergey Borisenko}
\affiliation{IFW Dresden, P.O. Box 270116, 01171 Dresden, Germany}
\author{Bernd B\"uchner}
\affiliation{IFW Dresden, P.O. Box 270116, 01171 Dresden, Germany}
\affiliation{Department of Physics, TU Dresden, 01062 Dresden, Germany}
\author{Jeroen van den Brink}
\affiliation{IFW Dresden, P.O. Box 270116, 01171 Dresden, Germany}
\affiliation{Department of Physics, TU Dresden, 01062 Dresden, Germany}
\affiliation{Department of Physics, Harvard University, Cambridge, Massachusetts 02138, USA}

\pacs{73.20.At, 71.55.Ak, 73.43.-f}

\title{TaIrTe$_{4}$ a ternary Type-II Weyl semi-metal}

\date{\today}
\begin{abstract}
{\jb
In metallic condensed matter systems two different types of Weyl fermions can in principle emerge, with either a vanishing (type-I) or with a finite (type-II) density of states at the Weyl node energy. So far only WTe$_{2}$ and MoTe$_{2}$ were predicted to be type-II Weyl semi-metals. Here we identify TaIrTe$_{4}$ as a third member of this family of topological semi-metals. TaIrTe$_{4}$ has the attractive feature that it hosts only four well-separated Weyl points, the minimum imposed by symmetry. Moreover, the resulting topological surface states -- Fermi arcs connecting Weyl nodes of opposite chirality -- extend to about 1/3 of the surface Brillouin zone.  This large momentum-space separation is very favorable for detecting the Fermi arcs spectroscopically and in transport experiments.
}
\end{abstract}
\maketitle

Weyl fermions are massless chiral fermions with a fundamental role in quantum field theory and high-energy physics\cite{Weyl_1929}.
In low-energy condensed matter systems Weyl fermions can emerge as quasiparticles: in semi-metals with either broken time-reversal symmetry or inversion symmetry, they can appear at the crossing point of two linearly dispersing three-dimensional bands\cite{Wan_2011,Weng_2015,Balents_2011}.
These Weyl nodes and their chirality are protected by the symmetry of the semi-metal and the topology of its electronic band structure.
As a consequence Weyl semi-metals (WSMs) host topologically protected surface states in the form of open arcs -- Fermi arcs -- terminating at the projection of bulk Weyl points (WPs) of opposite chirality.
Other physical consequences in WSMs are exotic transport phenomena such as a large negative magnetoresistance with unusual anisotropy\cite{Son_2013}.

Very recently it was pointed out that whereas in high-energy physics there is only one kind of Weyl fermion, in condensed matter systems there are two, and not more than two, types of WPs\cite{sol15}.
In a type-I WSM the bulk Fermi surface shrinks to a point at the Weyl node whereas in a type-II WSM the Weyl cone arises as a connector of  hole and electron pockets, where the large tilt of the cone induces a finite density of states at the node energy. This fundamental difference can for instance cause an anomalous Hall effect in type-II WSMs\cite{sol15,Zyuzin_2016,Ruan15}.
The type-I Weyl fermion is similar to the one also encountered in high-energy physics. In type-II systems, however, the Weyl nodes appear at touching points between electron and hole pockets so that the WP arising in this low-energy condensed matter setting\cite{Volovik14,YXu15} lacks a high-energy equivalent.

A number of type-I WSMs have been predicted theoretically \cite{Xu_2011,Wan_2011,Burkov_2011,Weng_2015,Huang_2015}. The first experimental indication for the presence of Weyl points and Fermi arcs has been established in a class of TaAs related compounds\cite{Xu_2015,Yang_2015,Lv_2015,SYXu15,DFXu15,Belopolski15,Souma15,NXu15,Chang15} using angle-resolved photoemission spectroscopy (ARPES) -- even if the large number of WPs (24) makes the interpretation of spectroscopic and transport data quite complicated in these TaAs-based systems. Type-II WSMs are very sparse at the moment -- this category has barely been explored and only two potential materials, orthorhombic WTe$_{2}$ and MoTe$_{2}$ have been identified theoretically\cite{sol15,Sun_2015,wang15}. Here we identify the orthorhombic ternary compound TaIrTe$_4$ as  a third type-II WSM. Based on calculations of its electronic structure and topological properties, we predict that TaIrTe$_4$ hosts only four type-II WPs, the minimal number of WPs a system with time-reversal invariance can host\cite{wang15}. The WPs are very well separated from each other in momentum-space: topological Fermi arcs that connect WPs of opposite chirality extend to about 1/3 of the surface Brillouin zone, and are present on both inequivalent surface terminations.  Such a large momentum-space separation is conducive for easy detection of the Fermi arcs in spectroscopy experiments and as well as to resolve their contribution to transport.

We performed density functional (DFT) calculations within the local
density approximation\cite{per92} using the full-potential
local-orbital code (FPLO)\cite{koe99}. For the Brillouin zone (BZ)
integration we used the tetrahdron method with a $12\times6\times6$
$\mathbf{k}$-mesh. Scalar relativistic calculations (without spin orbit
coupling (SOC)) and full 4-component relativistic calculations
(including SOC) were performed.  For easier analysis, a model has been
extracted using maximally projected Wannier functions (WFs) for the Ta
$5d$, Ir $5d$ and Te $5p$ orbitals.  The resulting (high accuracy)
model reproduces the band energies of all occupied bands and
unoccupied bands up to 1 eV above the Fermi level accurately. For
faster evaluation of all subsequent calculations the real space extent
of the WFs has been reduced resulting in a model which shows some
deviations of up to 10 meV around and slightly below the Fermi level,
especially off the $k_{z}=0$ plane. These deviations of the reduced
model slightly change the Fermi surface (FS) especially for $E=0$ eV
but not in the important BZ region. We have checked that all
topological properties are present in the accurate WF model as well.

TaIrTe$_{4}$ crystallizes in a non-centrosymmetric orthorhombic
structure with space group 31 (Pmn$2_{1}$)\cite{mar92}. The lattice
constants are $a=3.77$ \AA{}, $b=12.421$ \AA{} and $c=13.184$
\AA{}. The unit cell contains 4 formula units which form two layers of
distorted corner sharing octahedra. This compound has the same space
group as MoTe$_{2}$ and WTe$_{2}$ but in contrast to those, the size
of the unit cell is doubled in the $b$-direction
(Fig. \ref{fig:Crystal-structure-bands}a).  A detailed list of the
Wyckoff positions can be found in the supplementary
materials.

\begin{figure}[t]
\noindent \begin{centering}
\includegraphics[width=0.95\columnwidth]{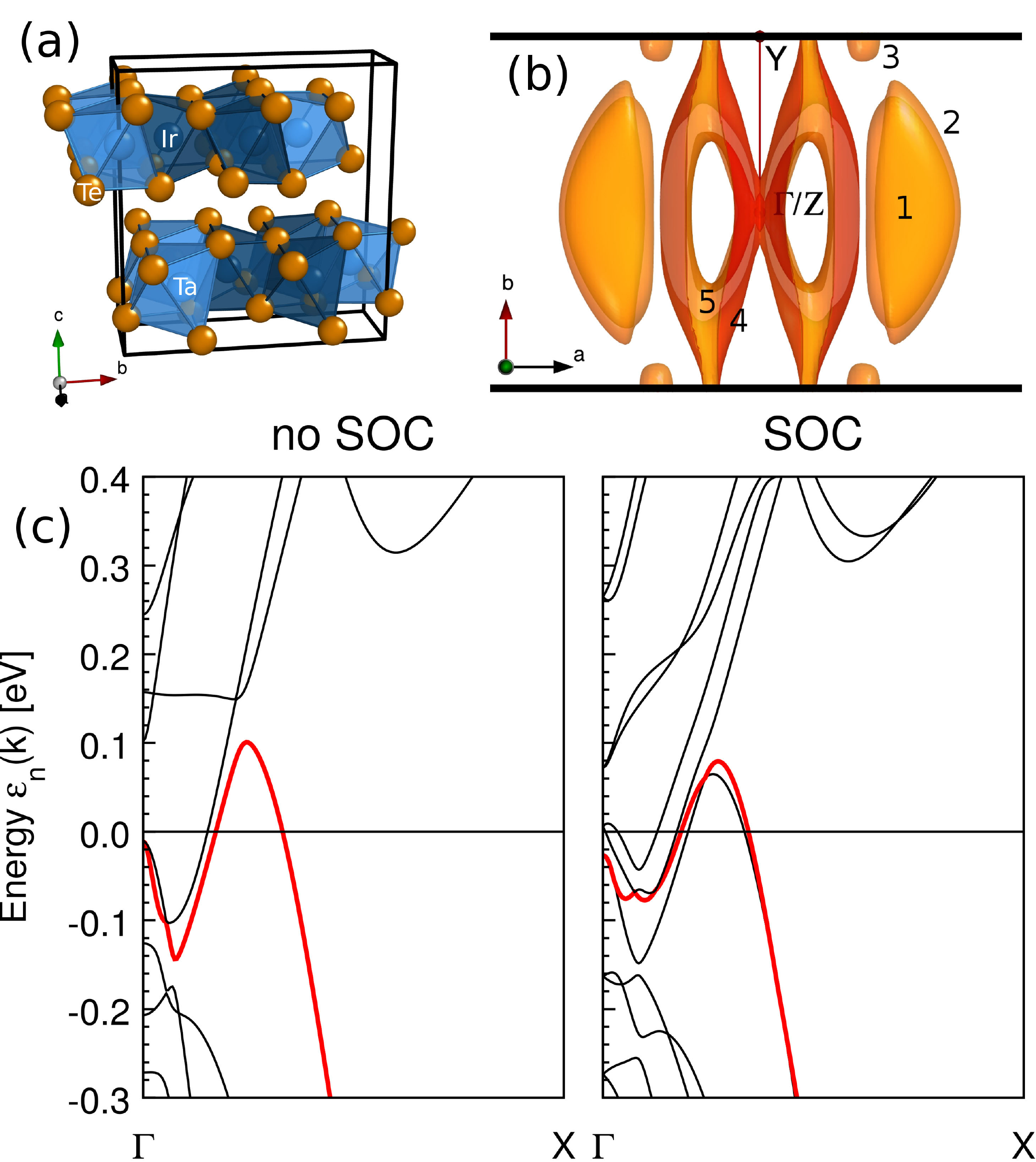}
\par\end{centering}

\caption{\label{fig:Crystal-structure-bands}(color online) (a) Crystal
  structure, (b) Fermi surface of the accurate WF model and (c) band
  structure along the line $\Gamma\mbox{X}$ both without and with
  SOC. Only the part of the BZ containing Fermi surfaces is shown in
  panel (b).}
\end{figure}

\begin{table}[b]
\begin{ruledtabular}
\begin{tabular}{lcccrr}
 &Band $N$& SOC & Position  & $C$ & $E$ [meV] \\
 &        &     &    in $(\frac{2\pi}{a},\frac{2\pi}{b},\frac{2\pi}{c})$ & & \\ \hline
WP1  &60& no  & (0.1178, 0.1529, 0) & -1 & 115\\
WP2  &60& no  & (0.0483, 0.2306, 0) & +1 & -79\\ \hline
WP  &120& yes & (0.1237, 0.1627, 0) & -1 & 82.7\\
\end{tabular}
\end{ruledtabular}

\caption{\label{tab:chern-numbers-table}Positions, Chern numbers and energies
of the Weyl points formed by the topmost valence band $N$ and the
lowest conduction band $N+1$.}

\end{table}

Although sharing the same space group with the binary tellurides, the
similarities of the TaIrTe$_4$ bandstructure compared to MoTe$_{2}$
and WTe$_{2}$ should not be overemphazised, since the Fermi surfaces
are quite different. Figure \ref{fig:Crystal-structure-bands}c shows
the band structure along the line $\Gamma\mbox{X}$ without and with
SOC. The band shown in red forms the top of the valence band. It
gives rise to three dimensional hole pockets along the line
$\Gamma\mbox{X}$. Upon inclusion of SOC, all bands split almost
everywhere, which leads to two nested but not touching hole pockets
(Fig. \ref{fig:Crystal-structure-bands}b, labeled 1 and 2). The
conduction bands around the $\Gamma$-point form two electron pockets
(labeled 4 and 5) which exhibit a noticeable two-dimensional character
in $k_{z}$-direction in contrast to the hole pockets. In contrast to
MoTe$_2$ and WTe$_2$, the hole and electron character of the pockets
are interchanged in TaIrTe$_4$. Note, that pockets 4 and 5 have open
Fermi surfaces across the $k_{y}=\pm\pi$ BZ boundaries for
$k_{z}=\pm\pi$. This feature is absent in the reduced WF model at
$E=0$ eV but reemerges $8$ meV above the Fermi level. Additionally,
small hole pockets (Fig. \ref{fig:Crystal-structure-bands}b, label 3)
occur at the $k_{y}=\pm\pi$ planes. Another notable difference to
MoTe$_2$ and WTe$_2$ is the fact that the top of the valence band
crosses the bottom of the conduction band close to the $\Gamma$-point
along the line $\Gamma\mbox{X}$ in TaIrTe$_4$.

\begin{figure}[t]
\noindent \begin{centering}
\includegraphics[width=0.95\columnwidth]{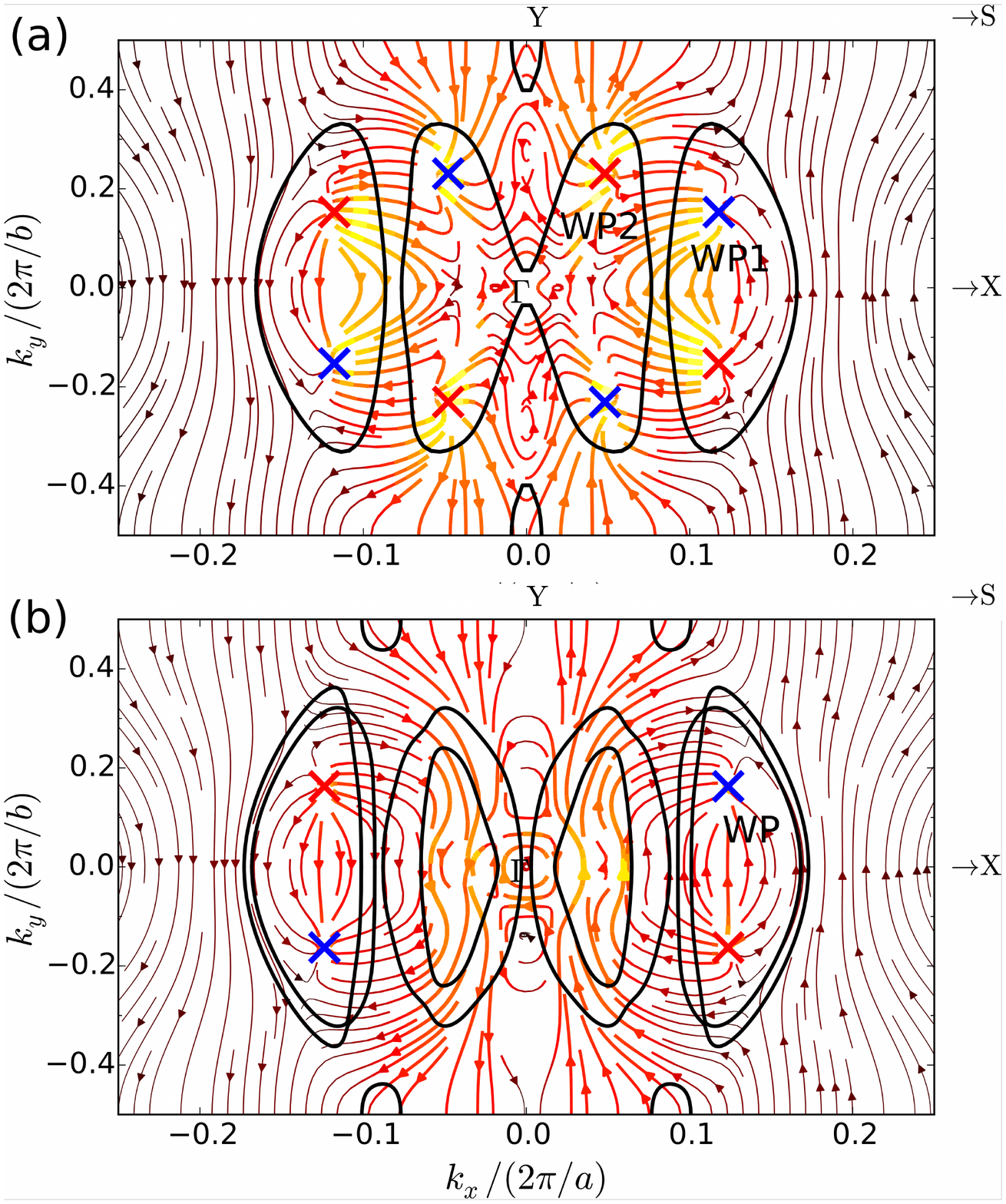}
\par\end{centering}

\caption{\label{fig:berryflux}(color online) The flux of the
  $k_{z}$-averaged Berry curvature in the $k_{x},k_{y}$-plane (a)
  without and (b) with SOC. Crosses mark the Weyl points (blue $C=-1$,
  red $C=+1$). The black solid curves are $k_{z}=0$-cuts through the
  bulk Fermi surface at $E=0$ eV.}
\end{figure}

Despite the difference in the hole/electron nature of the pockets,
TaIrTe$_4$ has similarities in its topological properties compared to
the two binary tellurides. Due to symmetry conditions Weyl points can only
occur in the $k_{z}=0$ plane\cite{sol15,wang15}. Without SOC we find 8
(spinless) Weyl points between the topmost valence band $N$ and the
lowest conduction band $N+1$. Each quadrant of the $k_{x}$,
$k_{y}$-plane contains two such points with opposite chern number
$C$. The one with negative chirality (WP1) recides in the hole pocket
around $k_{x}=0.1178$, while the one with positive chirality (WP2)
resides in the electron pocket around $k_{x}=0.0483$ (see Table
\ref{tab:chern-numbers-table} and Fig. \ref{fig:berryflux}).

With SOC, the spin doubled Weyl points get split and moved around and
due to the strength of the SOC, pairs of WP2 with opposite chirality
annihilate in the process. Of the spin doubled Weyl points WP1 only a
single point (WP) survives per quadrant which results in a total of 4
Weyl points in the BZ if SOC is included. Two WP with opposite
chirality sit in the hole pocket around $k_{x}=0.1237$ resulting in
zero chern number for the hole pocket. It is interesting to note that
WP1 and WP2 (without SOC) are type I Weyl points (see supplementary
materials) while WP (with SOC) is type-II.

In order to establish the topology of the electronic structure, we
implemented the calculation of the Berry curvature
$\mathbf{F}\left(\mathbf{k}\right)=\nabla_{\mathbf{k}}\times\mathbf{A}\left(\mathbf{k}\right)$
with $\mathbf{A}\left(\mathbf{k}\right)=-i\left\langle
u_{\mathbf{k}}\mid\nabla_{\mathbf{k}}u_{\mathbf{k}}\right\rangle $
into FPLO following the method of Ref. \onlinecite{wang06}. From this
we calculated the Chern number
$C=\frac{1}{2\pi}\oint_{s}\mathbf{F}\cdot d\mathbf{S}$ as the flux of
the Berry curvature of all bands upto and including band $N$ through a
closed surface surrounding the Weyl point. Here, we applied the
technique of choosing a small sphere around the WP such that bands $N$
and $N+1$ are separated by a gap all over the
sphere\cite{sol15,wang15}.  The results are summarized in Table
\ref{tab:chern-numbers-table}.

For illustration, the Berry curvature of the $N$ lowest bands was
calculated over the whole BZ followed by an integration over $k_{z}$
which results in the averaged Berry curvature projected onto the $k_{x}$,
$k_{y}$-plane. The resulting flux picture is shown in Fig. \ref{fig:berryflux}
without SOC (a) and with SOC (b) (Note, that only half the extent
of the BZ in $k_{x}$-direction is shown). With the sign choice of
the Berry connection, the sinks of the flux have $C=-1$ and flux sources
have $C=1$. The monopoles at the Weyl points are clearly visible
and lay well within the $k_z=0$-cuts of the $E=0$ bulk FS (solid lines). It
is also obvious that there are no other Weyl points between bands
$N$ and $N+1$.

There certainly are additional topological features in the bandstructure
such as line nodes on the $k_{x}$, $k_{z}$-plane as well as Weyl
points between different sets of bands. We will not discuss them in
this work mainly due to the fact that they will either be at energies
further away form the Fermi energy or fall into the bulk band projections.

\begin{figure*}[t]
\begin{centering}
\includegraphics[width=0.99\textwidth]{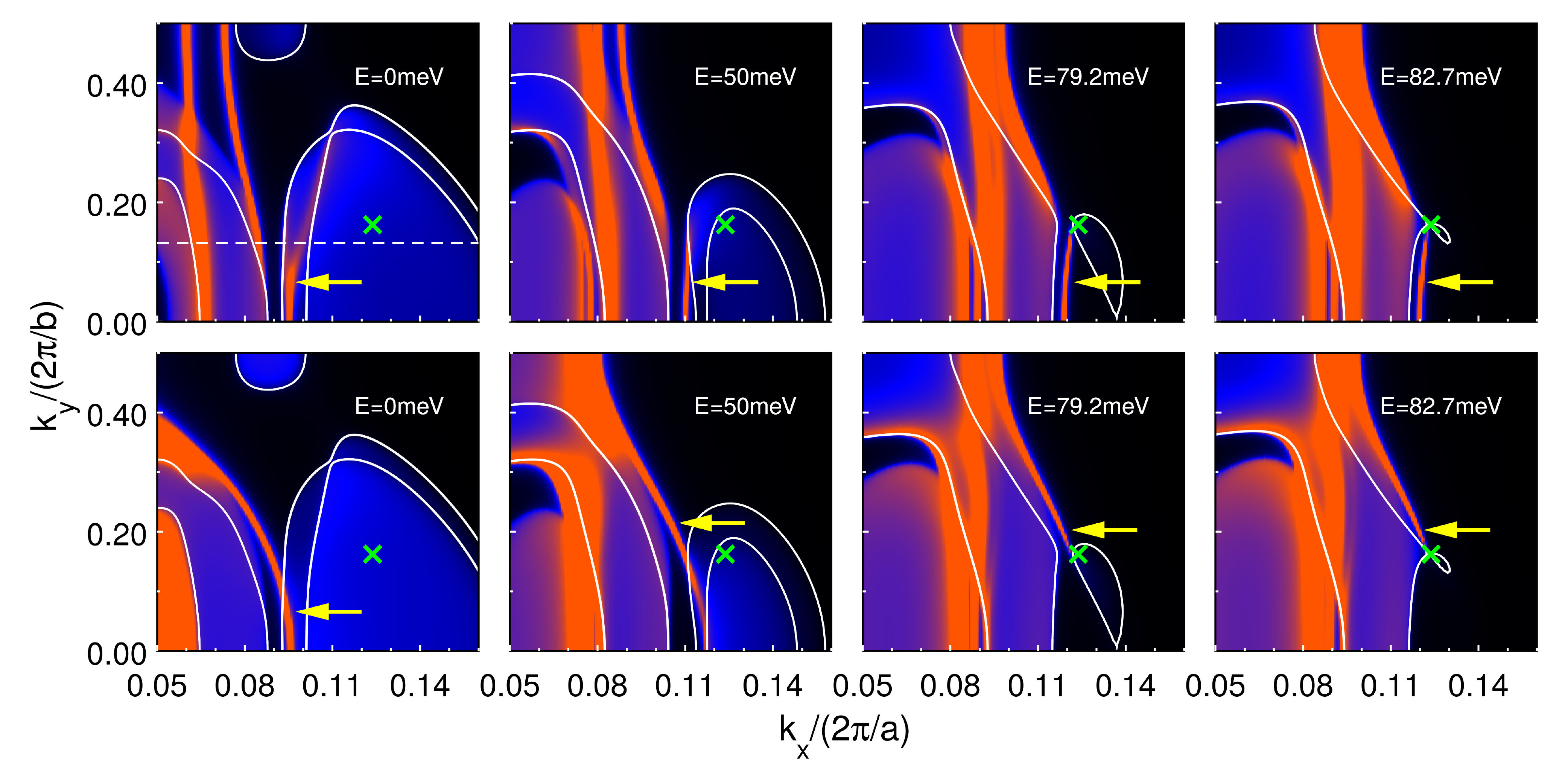}
\par\end{centering}

\caption{\label{fig:surface-FS}(color online) Surface Fermi surface with SOC
of (upper row) the $\left(001\right)$-surface and (lower row) the
$\left(00\overline{1}\right)$-surface for various energies. The position
of the WP is marked with a cross. The emergence of the Fermi arcs are is marked by
the arrows}. The solid lines are $k_z=0$ cuts
of the $E=0$ bulk Fermi surface.
\end{figure*}

{\jb
The Weyl points identified above give rise to topologically protected surface states.
}
In our case the projection of the Weyl points onto the
$\left(001\right)$-direction, map single distinct WPs onto points in
the surface BZ, in which case topologically protected Fermi arcs are
to be expected\cite{sol15}.  Figure \ref{fig:surface-FS} shows the
(001)-surface Fermi surface (spectral function) of an ideal semi
infinite slab. The spectral function was obtained via Green function
recursion \cite{san84,san85} by mapping the bulk WF Hamiltonian onto
the slab geometry while neglecting any surface and relaxation induced
modifications to the Hamiltonian. Only part of the surface BZ is shown
to focus on the essential details. Depicted is the spectral density of
the first unit cell at the surface, i.e. the two topmost octahedral
layers. For orientation, the Weyl point is marked by a cross and
$k_{z}=0$-cuts of the $E=0$ bulk FS are shown as solid lines.

A non-centrosymmetric compound has two distinct surfaces:
$\left(001\right)$ and $\left(00\overline{1}\right)$. Any Fermi arc
emerging from a WP must be accompanied by another different arc living
on the opposite surface. We choose two such surfaces by chosing the
natural cleavage plane between the octahedral layers followed by a
removal of either the upper or lower semi infinite half of the
bulk. The upper and lower panel of Fig. \ref{fig:surface-FS} show the
evolution of the spectral function of the $\left(001\right)$- and
$\left(00\overline{1}\right)$-surface, respectively, while increasing
the energy from the bulk Fermi level at $0$ meV to the WP energy
$82.7$ meV. At $0$ meV the spectral function of the
$\left(001\right)$-surface shows the two hole pockets 1 and 2
(Fig. \ref{fig:Crystal-structure-bands}), containing the WP, as
well as the hole pocket 3 and electron pockets 4 and 5 (around
$k_{x}=0.05$). Additionally, two trivial SO-split surface states (SS)
emerge out of the electron pockets running parallel to $k_{y}$ at
$k_{x}\approx0.07$. The high intensity feature at the left side of the
hole pockets marked by the arrow is the precursor of the Fermi arc,
which will develop with increasing energy. At $0$ meV, it is
submerged into the bulk projected bands (BPB). At $50$ meV the
electron pockets form open FSs along the $k_{y}$-direction while the
hole pockets shrink to such an extent that the Fermi arc is separated
from the BPB forming a true SS. Yet, still the hole pockets are open
across the $k_{y}=0$ plane and contain two WP with total Chern number
zero, which opens the possibility of the arc being absorbed into the
BPB due to perturbations. There is however no prohibition for an arc
to be attached to a FS with zero total chern number\cite{hal14}.

This changes at a critical energy ($E=79.2$ meV) at which hole
pocket 1 is collapsed and pocket 2 gets separated into two distinct
closed pockets containing WPs of opposite chirality. At this stage
the Fermi arc should be protected against perturbations. If the energy
is raised to the WP energy, hole pocket 2 remains of finite size and
has connected to the electron pocket at the coordinates of the Weyl
point, which shows the WP to be type-II. Note, how the Fermi arc is
following the flux lines of the Berry curvature (Fig. \ref{fig:berryflux}b).

The lower row of Fig. \ref{fig:surface-FS} depicts the
$\left(00\overline{1}\right)$-surface FS for the same set of
energies. The noticeable differences with respect to the other surface
are the absence of the $k_{y}$-parallel SS and the emergence of the
state connecting the hole pockets with the electron pockets marked by
the arrow. This state evolves with increasing energy such that the
point where it connects to the hole pocket moves closer to the WP,
finally emerging exactly out of the Weyl point at WP energy.  At
$E=79.2$ meV it can be best seen that this state is also a Fermi arc
only that it does not connect two Weyl points in the same BZ as the
arc on the other surface. It rather connects two such points in
adjacent BZs via immersion into the electron bulk states (see
supplementary materials for further elucidation.) At the WP energy it
is not a free standing arc anymore, due to it forming a surface
resonance at the edge of the electron pocket.
We checked thatthese arc features even survive the enormous perturbation
introduced by cleaving the cell in the middle of the octahedral layer
such that the surface is terminated by an unsaturated transition metal
layer. A slew of bond breaking related surface states
appear in such a scenario. Yet the arcs remain, especially at energies
between $50$ and $82.7$ meV.

\begin{figure}[b]
\begin{centering}
\includegraphics[width=0.95\columnwidth]{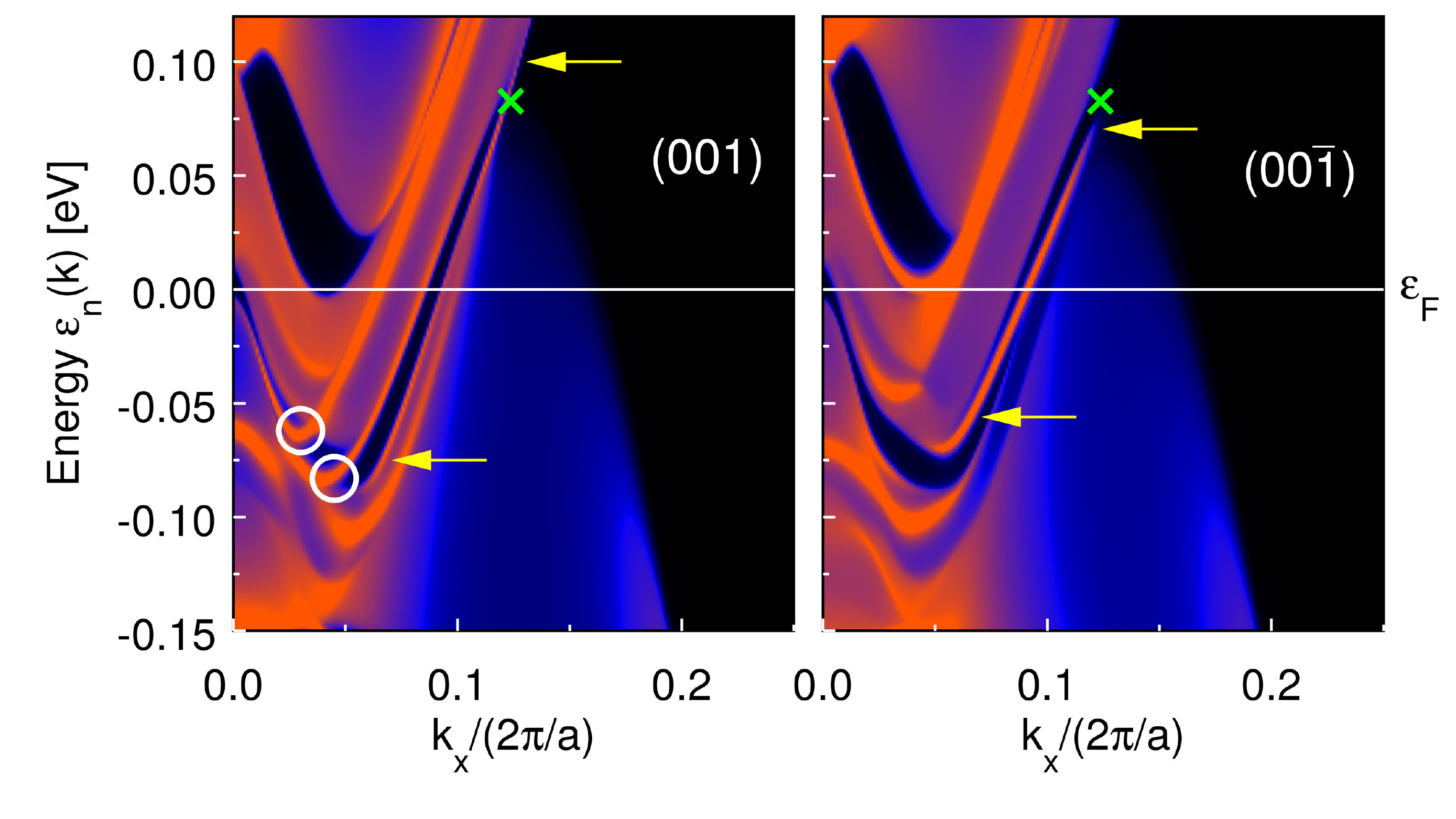}
\par\end{centering}

\caption{\label{fig:Surface-EDC}Surface energy density cuts along the line
$\mathbf{k}=\left(\frac{2\pi}{a}k_{x},\frac{2\pi}{b}\cdot0.132,0\right)$
for the two surfaces.}

\end{figure}

To further elucidate the electronic structure of the Fermi arcs, we
show the energy density cuts of the spectral density of the topmost
unit cell of the semi infinite slab for the $\left(001\right)$- and
$\left(00\overline{1}\right)$-surfaces along the line $\mathbf{k}=\left(\frac{2\pi}{a}k_{x},\frac{2\pi}{b}\cdot0.132,0\right)$
in Fig. \ref{fig:Surface-EDC}. The cut line is indicated by the
dashed line in Fig. \ref{fig:surface-FS}. The Fermi arc of the $\left(001\right)$-surface
is formed by a band which connects the top of valence bands around
$k_{x}=0.07$ to the flank of the conduction band bottom around $k_{x}=0.13$
as indicated by the arrows in the left panel of Fig. \ref{fig:Surface-EDC}.
The two intense bands in the electron pocket marked by circles are
the two trivial SS visible for $E=0$ meV in the upper row of
Fig. \ref{fig:surface-FS}. The band forming the Fermi arc of the
$\left(00\overline{1}\right)$-surface connects the conduction band
minimum around $k_{x}=0.05$ to the top of the valence band of the
hole pocket around $k_{x}=0.12$. Hence, if superimposed in the same
picture the two surface bands of the opposite surfaces represent two
crossing bands bridging the gap as further illustrated in the supplementary
material.

In summary we have demonstrated via DFT methods that the ternary
compound TaIrTe$_{4}$ is a type-II Weyl semi-metal with the simplest
possible arrangement of Weyl points under the symmetry constraints of
the compound.  Topological Fermi arcs are present on both of the
surfaces of a naturally cleaved ideal semi infinite slab, as well as
for
{\jb
other
}
surface terminations. The length of the emerging Fermi
arc of the $\left(001\right)$-surface is about $1/3$ of the BZ extent
in $b$-direction. The energy range in which the Fermi arcs should be
detectable reaches from $50$ meV to $82$ meV above the Fermi
level. All three telluride compounds show at least the minimum set of
4 Weyl points, occurring at very similar positions in $k$-space.  This
seems to indicate that the existence of these particular Weyl points
are a generic feature in this family of compounds, and it is therefore likely that
slight modifications of the band filling ($i.e.$ doping) can be used
to bring the the bulk Weyl points and the surface arcs connecting them at or even below
the Fermi level where ARPES can resolve them.

\begin{acknowledgments}
K. Koepernik is grateful to Hongbin Zhang for valuable information
exchange. D. Kasinathan acknowledges funding from the Deutsche Forschungsgemeinschaft  (DFG) via FOR1346.
This work was supported by the DFG through the Collaborative Research Center SFB 1143. JvdB acknowledges support from the Harvard-MIT CUA. 
\end{acknowledgments}

\end{document}